\documentclass[12pt,twoside]{article}
\usepackage{xrb2007}
\usepackage{epsf}
\usepackage{graphics}

\begin{document}

\session{Obscured XRBs and INTEGRAL Sources}

\shortauthor{Negueruela et al.}
\shorttitle{SFXTs and wind accretion}

\title{Supergiant Fast X-ray Transients and Other Wind Accretors}
\author{Ignacio Negueruela$^{1}$, Jos\'e~Miguel~Torrej\'on$^{1,2}$, Pablo Reig$^{3}$, Marc Rib\'o$^{4}$, David M.~Smith$^{5}$}
\affil{$^{1}$DFISTS, EPSA, Universidad de Alicante, Apdo. 99,
  03080 Alicante, Spain}
\affil{$^{2}$M.I.T., Kavli Institute for Astrophysics
and Space Research, Cambridge MA 02139}
\affil{$^{3}$IESL (FORTH) \& Physics Dpt, University of Crete, 71003, 
                Heraklion, Greece}
\affil{$^{4}$ DAM,
Universitat de Barcelona,
Mart\'{\i} i Franqu\`{e}s 1, 08028 Barcelona, Spain}
\affil{$^{5}$University of California Santa Cruz, 1156 High St., Santa
  Cruz, CA 95064}

\begin{abstract}
Supergiant Fast X-ray Transients are obviously related to
persistent Supergiant X-ray Binaries. Any convincing explanation
for their behaviour must consistently take into account all
types of X-ray sources powered by wind accretion. Here we present a common
framework for wind accreting sources, within the context of clumpy
wind models, that allows a coherent interpretation of their different
behaviours as an immediate consequence of
     diverse orbital geometries. 
\end{abstract}

\section{Supergiant Fast X-ray Transients}

Traditionally, High Mass X-ray binaries (HMXBs) have been divided in
three classes: ``classical'' bright ($L_{{\rm X}}\sim10^{38}\:{\rm
  erg}\,{\rm s}^{-1}$) sources fed by incipient localised Roche lobe
overflow, Be/X-ray binaries, presenting different behaviours but always
associated to the presence of a circumstellar decretion disk, and
wind-fed Supergiant X-ray Binaries (SGXBs). In the last group, the
compact object is fed by accretion from the strong radiative wind of
an OB supergiant, leading to a persistent X-ray source with $L_{{\rm
    X}}\sim10^{36}\:{\rm erg}\,{\rm s}^{-1}$, displaying large
variations on short timescales, but a rather stable flux in the long
run. The compact object is a relatively close ($P_{{\rm
    orb}}\la15\:{{\rm d}}$) orbit with low or negligible
eccentricity.

Over the last few years, there has been a huge increase in the number
of HMXBs known \citep[e.g.,][]{wal06}.
Many new sources have been found to display very brief outbursts, with a
rise timescale of tens of minutes and lasting only a few hours
\citep{sgue05}. Several have been identified with OB supergiants:
 \citep[e.g., the prototype XTE~J1739$-$302;][]{smith06,neg06b}. The
 distances to their counterparts imply typical $L_{{\rm
    X}}\sim10^{36}\:{\rm erg}\,{\rm s}^{-1}$ at the peak of the
outbursts. A number of other systems have been shown to have similar
X-ray behaviours leading to the
definition of a class of Supergiant Fast X-ray Transients (SFXTs)
\citep{esa,smith06}. 

Around 12 such systems are known \citep{wzh07}. It is perhaps possible
to divide them in two groups, the first one characterised by
very low quiescence $L_{{\rm X}}$ and high variability and the second
one with higher average $L_{{\rm X}}$ and smaller variability
factors. The second group, then, could be thought of as persistent
SGXBs with an average $L_{{\rm X}}$ below the 
  canonical $\sim10^{36}\:{\rm erg}\,{\rm s}^{-1}$ over which flares
  are superimposed \citep{wzh07}. To some degree, this distinction may
  be due to observational biases (cf. the case of
    XTE~J1739$-$302; Blay et al., submitted). In any case, it is clear that the
  separation between SFXTs and SGXBs is not well defined.

Two other systems have shown flaring only
 during episodes separated by regular intervals: IGR J00370+6122
 \citep{zand07} and
 IGR J11215-5952 \citep{sidoli}. This modulation is most likely
 related to an orbital period. Because of their very different
 properties, we do not count these two systems as
 SFXTs, but they are also
 wind-fed accretors and any theory attempting an explanation to the behaviour
 of SFXTs must take into account the existence of other wind
 accretors.

At least two SFXTs have been observed to increase their $L_{{\rm
    X}}$ by a factor $>100$ (going from deep quiescence to outburst)
in only a few minutes: XTE~J1739$-$302 \citep{sak02} and
IGR~J17544$-$2619 \citep{zand05}.  AX~J1841.0$-$0535 showed a $>10$
increase in count-rate over $\la 1\:$h \citep{bam01}. Such sharp rises
are incompatible with any explanation of their behaviour based solely
on orbital motion through a smooth medium.

\section{Wind accretion, clumps and X-ray bahaviour}
Radiation-driven winds from hot stars obey complex physics
\citep{kp00}. The outflow is driven by line scattering of the
continuum radiation flux from the central star. Material is
accelerated outwards from the stellar atmosphere to a final velocity
$v_{\infty}$ according to a law that may be approximated as
\begin{equation}
v_{\rm w}(r)=v_{\infty}\left(1-\frac{R_{*}}{r}\right)^\beta
\, ,
\end{equation}
where $R_{*}$ is the radius of the supergiant and $\beta$ is a factor
generally lying in the interval $\sim0.8-1.2$. 
The accretion rate ($\dot{M}$) of a neutron star immersed in such a
wind depends on its accretion radius, the maximum distance at which its
gravitational potential well can deflect the stellar wind and focus
the outflowing material towards the neutron star. On first
approximation, this is given by
\begin{equation}
 r_{{\rm acc}} \sim 2GM_{{\rm X}}/v^{2}_{{\rm rel}}
\, ,
\end{equation}
where the relative velocity of the accreted
material with respect to the neutron star is $v^{2}_{{\rm
    rel}}= v^{2}_{\rm w} + v^{2}_{{\rm orb}}$. For typical values of
$v_{\rm w}$,  $r_{{\rm acc}}\sim 10^{8}\:{\rm m}$, decreasing by a factor of a
few from $r = 2\:R_{*}$ to $r = 10\:R_{*}$. Under these conditions, 
$L_{{\rm X}}\sim \dot{M} \sim \rho(r)v_{\rm w}(r)/v^{4}_{{\rm rel}}
\, .$

The details of wind accretion are not well
understood. In the classical
Bondi-Hoyle formulation, accreted material lacks
enough angular momentum to form a
stable accretion disk that can provide torque to the compact
object. Neutron stars in many SGXBs alternate spin-up and spin-down episodes,
resembling a random walk \citep{ab95}. Complex
simulations of the accretion process \citep[][and references
  therein]{ruff99} are unable to quantify how much angular momentum is
accreted from a wind. 

Direct application of this formulation implies that SGXBs must display a
wide range of X-ray behaviours and $L_{{\rm X}}$, depending on orbital
characteristics. In contrast, all known SGXBs have about the same
$L_{{\rm X}}$. Moreover, all SGXBs known have their compact objects
orbiting the supergiant at $r\la2\,R_{*}$. 
The actual situation in real SGXBs is much more complex than the toy
model represented by these
simple equations  
\citep[e.g.,][]{blondin}, but the lack of a distribution in $L_{{\rm
    X}}$ amongst SGXBs is still unexpected if winds are smooth.


Because of its very nature, radiation-driven winds from hot stars must
be highly unstable to small-scale perturbations. Simulations
suggest that irregular variations in both
density and velocity develop at a small
height above the photosphere, leading to the appearance of structure
within the wind \citep[e.g.,][]{ro02}. The
variations steepen into shocks that decelerate and compress rarefied
gas, collecting most of the mass into a sequence of dense clumps
bounded by shocks. Indirect observational evidence supports the
existence of clumps within the wind, though the actual physical
parameters (size, geometry, density) of the clumps are ill-determined
\citep{do03}. Clumpiness may provide an explanation for the behaviour
of SFXTs, as the wind would be composed of dense clumps (with a density
contrast with respect to the corresponding
smooth wind model $D\la10$) separated by effectively void space. 

\begin{figure}
   \centering
  \resizebox{9cm}{!}{\includegraphics{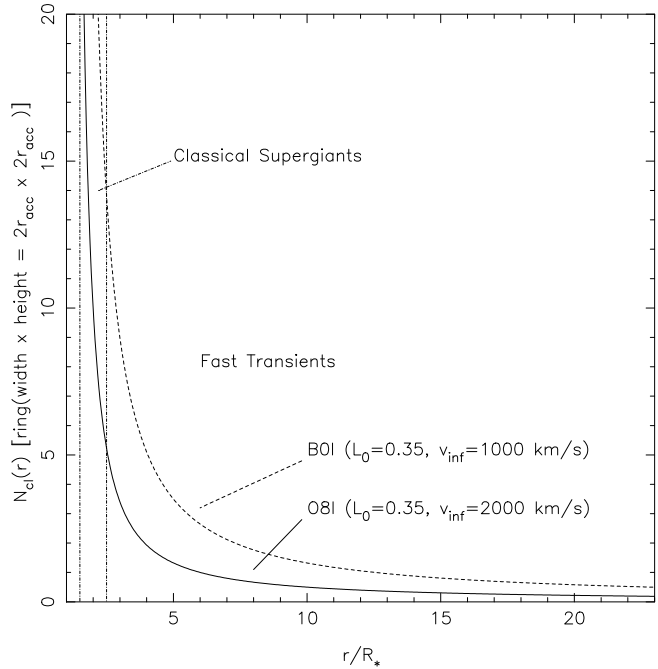}} 
      \caption{Number of clumps in a ring of width $2r_{{\rm acc}}$ and
        height $2r_{{\rm acc}}$ as a function of distance from the center of
        the star, using the model of \citet{oski07}. The distance
        is expressed in units of stellar radius ($r/R_{*}=1$ is the
        stellar surface). A velocity law with
        $\beta=0.8$ and a porosity parameter 
        $L_{0}=0.35$ have been assumed (but the shape of the graph does
        not depend on $L_{0}$). SGXBs can only lie within the two
        vertical dot-dashed lines, where a neutron star will see a
        very high number of clumps.}
         \label{clumps_distance}
   \end{figure}

We can identify each outburst from an SFXT as due to the accretion of a
single clump, assuming, like \citet{wzh07}, that the X-ray lightcurve
is a direct tracer of the density distribution
in the wind. Alternatively, if the
accretion process is very complex, a number of physical mechanisms may
mediate, such as the centrifugal effect of a neutron's star magnetic
field, as suggested by \citet{greb07}. In this case, $L_{{\rm X}}$ is
not a direct tracer of $\rho$. Our current understanding of wind accretion
supports the first scenario, as accretion disks seem difficult to form
and the $P_{{\rm orb}}/P_{{\rm spin}}$ distribution of SGXBs does not
suggest an effective transfer of angular momentum from wind to
accreting object \citep{wvk89}. Therefore we will investigate the wind distribution
that would be directly traced by the observed $L_{{\rm X}}$.

We calculate the clump distribution for the inner wind within the model of
\citet{oski07}, though we do not expect the choice of model to be
determinant.
In this formulation, the number of clumps within a spherical shell of width
$\Delta r$ is given by
\begin{equation}
\label{clumpsinshell}
N_{\rm cl}=\frac{4\pi}{L^{3}_{0}}\Delta t
\, ,
\end{equation}
where $\Delta t$ is the flight time that the wind takes to go through
$\Delta r$, expressed in units of the dynamical timescale
$R_{*}/v_{\infty}$, and $L_{0}$ is the porosity length.

We consider that a clump can be accreted if it is at a
distance $\leq r_{\rm acc}$ from the compact object. For typical sizes
of OB supergiants, $r_{{\rm acc}}\ll R_{*}$, and we can approximate the 
flight time as
$\Delta t\approx 2r_{{\rm acc}}^{*}/w(r)
\, ,$
where $r_{{\rm acc}}^{*}=r_{{\rm acc}}/R_{*}$.
Now we distribute the clumps uniformly over the shell
to produce a volume density of clumps
$\rho(r)=N_{{\rm cl}}^{{\rm shell}}/V^{{\rm shell}}$ (we assume the shell to be
spherical). Finally, we compute the number of clumps within a ring of
height $2r_{{\rm acc}}$ inside this shell (itself of width $2r_{{\rm acc}}$) as
$N_{{\rm cl}}^{{\rm ring}}(r)=\rho V^{{\rm ring}}$. $N_{{\rm cl}}$ gives an
indication of the number of clumps that the neutron star is {\it
  statistically} able to accrete in one orbit. Values of $N_{{\rm cl}}$ are
plotted in 
Fig.~\ref{clumps_distance} for typical parameters of O and B 
supergiants.

The \emph{shape} of the distribution, which does not depend on
$L_{0}$ and only weakly on wind parameters (within a sensible range)
suffers a change in the 
slope at a distance $r\sim 2\:R_{*}$. This change is so abrupt that 
 it effectively defines a two-regime scenario. At this distance, the
 distribution tends to a \emph{vertical asymptote}, which
 separates a region where the neutron star sees a very large number of
 clumps in the wind from a region where the number density of clumps
 is so small that the neutron star is effectively in empty
 space. Systems in which the neutron star orbits the supergiant at
 smaller distances than this asymptote are effectively embedded in a
 quasi-continuous wind. Systems outside the discontinuity spend most
 of the time in an effective vacuum. On this side, the number of clumps that 
 the neutron star encounters in one orbit must be very small.

In terms of the model, the interpretation of this shape is simple. At
$r\sim3\,R_{*}$, the wind has reached a 
significant fraction of its final velocity. For smaller values of $r$,
$r_{\rm acc}$ becomes large, reflecting the decrease in $v_{\rm
  w}$ (and hence $v_{\rm rel}$). At the same time, close to the star,
the wind occupies a smaller volume. The asymptote reflects the fact
that, for sufficiently small $r$, $r_{{\rm acc}}$ becomes comparable
in size to the volume filled by the wind. In other words, the neutron
star becomes able to ``see'' a significant fraction of all the clumps
coming out from the star. Conversely, at large distances, the wind
fills a much larger volume and the neutron star only ``sees'' a very
small portion of this volume. Obviously, the asymptote is due to the
mathematical formulation of the problem in a simple approximation, but
it has a clear physical interpretation in terms of the capability of the
neutron star to accrete. In spite of its simplicity,
this basic model of a porous wind predicts a substantial
change in the properties of the wind ``as seen by an orbiting neutron
star'' at a distance $r\sim2\:R_{*}$, exactly where we stop seeing
persistent X-ray sources.

 \begin{figure}
   \centering
  \resizebox{\textwidth}{!}{\includegraphics{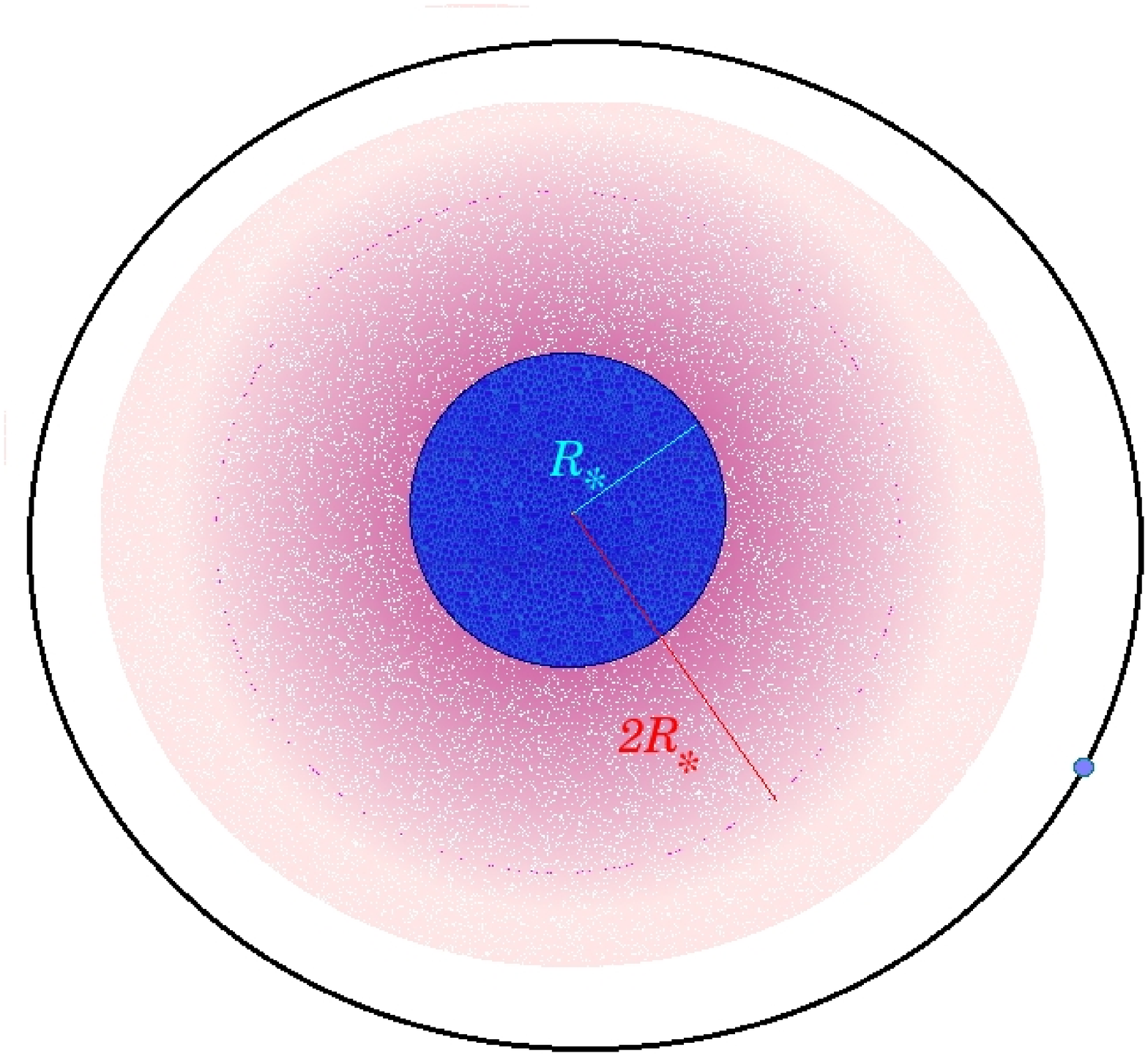}\includegraphics{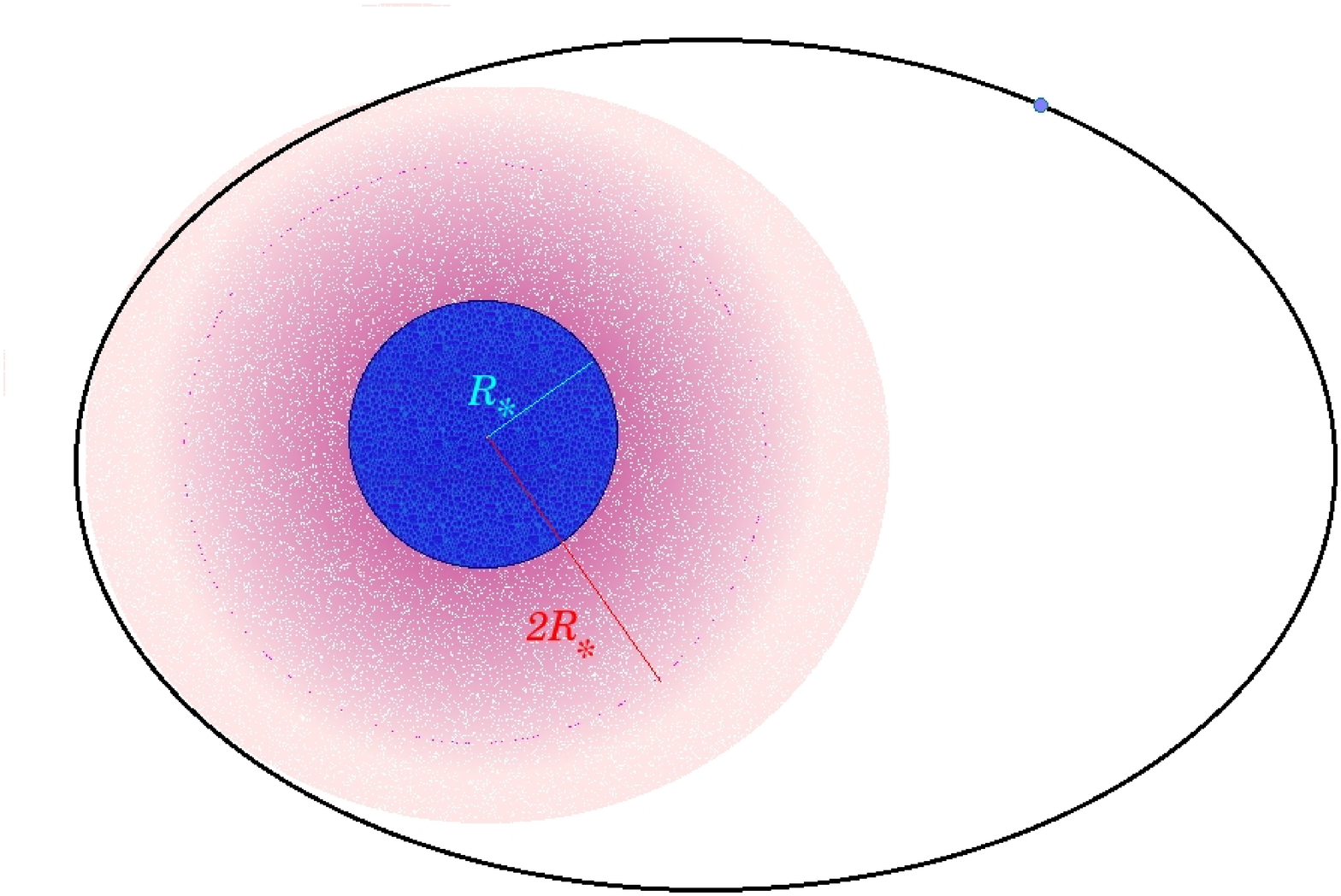}} 
      \caption{Two possible schematic configurations for an SFXT. The
        coloured area represents the region of the wind where the
        clump density is high (the area left of the asymptote in
        Fig.~1). The blank area represents the region 
        where the clump density is low. An SFXT may have an
        approximately circular orbit and lie just outside the high
        density area (it will be an ``intermediate'' system if it lies
        in the narrow transition zone) or a more eccentric orbit (it
        will tend to have longer quiescence intervals). For higher
        eccentricities, the neutron star may graze the coloured area
        at periastron (case of IGR J00370+6122). In a SGXB, the
        neutron star spends all its time inside the coloured area.
              } 
         \label{representation}
   \end{figure}

\section{Classes of wind accretors}

The scheme presented here not only provides a simple explanation for
the behaviour of SFXTs, but also the observed distributions of
SGXBs. Moreover, it naturally implies a smooth transition
between SGXBs and SFXTs and the existence of intermediate systems. As
a matter of fact, as shown in Fig.~\ref{representation}, all wind
accretors can smugly fit within this picture. 

The values of $L_{0}$ required to reproduce the observed outburst
frequency in SFXTs are relatively low ($0.1-0.3$). Such values are in
good agreement with those required by \citet{oski07} to fit UV line
profiles. Taken at face value, they imply a very high degree of
porosity (macroclumping). However, this is not the only possible
interpretation. A minimal mediation by the accretion process (for
example, if accretion is centrifugally inhibited for very low wind
densities) would probably allow the observed X-ray behaviour even with
rather smaller porosity parameters. 

Under the assumptions presented here, the observed division of hard
X-ray supergiant sources into persistent SGXBs, SFXTs and regular
outbursters can be naturally explained by simple geometrical
differences in their orbital configurations.

\acknowledgements This research is partially supported by the MEC under
grants AYA2005-00095 and AYA2007-68034-C03-01 and the European Union Marie Curie
grant MTKD-CT-2006-039965. MR holds a {\it Ram\'on y Cajal} fellowship.

\end{document}